\documentclass[a4paper,11pt]{article}
\pdfoutput=1
\usepackage{latexsym}
\usepackage{amsmath, amsthm, amssymb, bbm}
\usepackage[mathscr]{eucal}
\usepackage{cite}
\usepackage{hyperref}
\usepackage{graphicx}
\usepackage{fullpage}
\usepackage{xcolor}

\newcommand{\rhs}{r.h.s.\ }

\newcommand{\wrt}{w.r.t.\ }
\newcommand{\cf}{cf.\ }

\newcommand{\bra}[1]{\langle #1 \rvert}
\newcommand{\ket}[1]{\lvert #1 \rangle}
\newcommand{\braket}[2]{\langle #1 \vert #2 \rangle}

\newcommand{\ud}{\mathrm{d}}
\newcommand{\del}{\partial}

\newcommand{\betrag}[1]{{\lvert #1 \rvert}}

\newcommand{\Z}{\mathbb{Z}}

\newcommand{\order}{\mathcal{O}}

\newcommand{\rD}{{\mathrm{D}}}
\newcommand{\rN}{{\mathrm{N}}}

\newcommand{\eps}{\varepsilon}

\newcommand{\nn}{\nonumber}
\newcommand{\beq}{\begin{equation}}
\newcommand{\eeq}{\end{equation}}

\DeclareMathOperator{\sign}{sgn}

\begin{document}

\title{Vacuum polarization near boundaries}
\author{Jonathan Wernersson\thanks{jonathan2008@live.se}, Jochen Zahn\thanks{jochen.zahn@itp.uni-leipzig.de} \\ Institut f\"ur Theoretische Physik, Universit\"at Leipzig \\ Br\"uderstr.~16, 04103 Leipzig, Germany}

\date{\today}

\maketitle

\begin{abstract}
We study the effect of boundary conditions on vacuum polarization for charged scalar fields in two space-time dimensions. We find that both Dirichlet and Neumann boundary conditions lead to screening. In the Dirichlet case, the vacuum polarization charge density vanishes at the boundary, whereas it attains its maximum there for Neumann boundary conditions. At a critical field strength, the vacuum polarization diverges for Neumann boundary conditions, an effect due to the instability of the lowest energy mode in the presence of the external field.
\end{abstract}

\section{Introduction}

Historically, vacuum polarization was one of the first effects of quantum electrodynamics that were theoretically studied \cite{Uehling35}. On the practical side, it provides corrections to atomic energy levels, \cf the review \cite{MohrPlunienSoff98}. In recent years, there has been revived interest in the topic, not only in its dynamical version, the Schwinger effect \cite{Dunne2008}, but also in the context of advanced materials such as 
 topological insulators \cite{Tkachov}. The latter topic in particular sparks the interest in the effect of boundaries on vacuum polarization \cite{FialkovskyKurkovVassilevich}.  

Despite the relevance of the topic, little seems to be known about the effect of boundary conditions on vacuum polarization. To the best of our knowledge, the only work in which the effect of different boundary conditions in external electric fields is compared is \cite{AmbjornWolfram83}. The authors considered the (massless) charged scalar field in two space-time dimensions, confined to an interval and subject to a constant electric field. It was found that, at linear order in the external field, Dirichlet boundary conditions exhibit screening (with the maximal charge density near the boundaries) and Neumann boundary conditions anti-screening. These findings are rather surprising and counter-intuitive. One would expect vacuum polarization to be always screening. Furthermore, due to the repulsive (attractive) nature of Dirichlet (Neumann) boundary conditions, one would expect the screening to be stronger for Neumann boundary conditions.

The purpose of this note is to show that these surprising results are due to two major flaws in the calculation: First, a mode sum formula for the charge density is used, which, as pointed out in \cite{Current}, can not be derived from a manifestly gauge invariant renormalization scheme and leads to incorrect results. Second, the perturbative expansion used in \cite{AmbjornWolfram83} breaks down for the zero mode in the massless case with Neumann boundary conditions. Hence, this mode was neglected in \cite{AmbjornWolfram83}. However, the zero mode is the mode that is most sensitive to external fields and thus contributes most to vacuum polarization. Here, we use a corrected version of the mode sum formula and consider massive fields to avoid the problem with the zero mode in the Neumann case. We obtain the expected behavior of vacuum polarization: It is screening and more pronounced for Neumann than for Dirichlet boundary conditions. More precisely, we find that the charge density vanishes near the boundary in the Dirichlet case and attains its maximum at the boundary in the Neumann case. This also complies nicely with the finding that the current density vanishes near Dirichlet boundaries in an Aharonov-Bohm type setting with toroidally compactified dimensions \cite{BellucciSaharianSaharyan15}. 

These results are first derived perturbatively, at first order in $\lambda$. A non-perturbative analysis vindicates these results in the Dirichlet case. However, for Neumann boundary conditions it reveals a divergence of the vacuum polarization at a critical field strength. The divergence occurs at the field strength for which the lowest mode has vanishing frequency and becomes non-normalizable. Beyond the critical field strength, the frequency of this mode becomes imaginary, signalling an instability.

This article summarizes and extends the results of the B.Sc.\ thesis of J.W., written in 2015 at the Institute of Theoretical Physics, Leipzig University, under the supervision of J.Z.  

\section{Setup}

We follow the conventions of \cite{AmbjornWolfram83}, i.e., we use signature $(+, -)$, and define the covariant derivative as
\beq
\label{eq:D}
 D_\mu \phi = \del_\mu \phi + i e A_\mu \phi.
\eeq
We denote our coordinates by $x = (t, z)$. The field $\phi$ satisfies the Klein-Gordon equation
\beq
\label{eq:KG}
 (D_\mu D^\mu + m^2) \phi = 0
\eeq
and the corresponding charge density, i.e., the $0$ component of the current, is given by
\beq
 \rho = i e \left( \phi^* D_0 \phi - \phi (D_0 \phi)^* \right).
\eeq

We will be interested in the vacuum polarization on a finite interval, whose length we normalize to 1 so that $z \in [0,1]$, in the presence of a constant electric field $E$. The latter is implemented by
\begin{align}
\label{eq:A}
 A_0 & = - E ( z - \tfrac{1}{2} ), & A_1 & = 0. 
\end{align}
With the separation ansatz
\beq
 \phi = \phi_n(z) e^{- i \omega_n t},
\eeq
the equation of motion \eqref{eq:KG} can be explicitly solved in terms of parabolic cylinder functions as
\beq
\label{eq:ExactModeSolution}
 \phi_n(z) = a_n D_{i \frac{m^2}{2 \lambda} - \frac{1}{2}}( \tfrac{1+i}{\sqrt{\lambda}} ( \omega_n + \lambda (z - \tfrac{1}{2}) ) ) + b_n D_{-i \frac{m^2}{2 \lambda} - \frac{1}{2}}( \tfrac{i-1}{\sqrt{\lambda}} ( \omega_n + \lambda (z - \tfrac{1}{2}) ) ),
\eeq
where
\beq
 \lambda = e E.
\eeq
With these mode functions, the evaluation of the vacuum polarization has to proceed numerically. It is thus more instructive to proceed perturbatively, i.e., to consider the electric field as a perturbation and compute the vacuum polarization at first order in $\lambda$. For this, the corrections of first order in $\lambda$ of the frequencies and the mode functions have to be determined. For this purpose, it is useful to write, as in \cite{AmbjornWolfram83}, the mode frequencies $\omega_n$ and solutions $\phi_n$ as solutions to a time-dependent Schr\"odinger equation, albeit in a space of indefinite metric. 
Introducing
\beq
 \Psi = \begin{pmatrix} \phi \\ \pi^* \end{pmatrix},
\eeq
with $\pi^* = D_0 \phi$ the momentum conjugate to $\phi^*$, we can write the equation of motion in the form
\beq
 i \del_t \Psi = H \Psi
\eeq
with
\beq
 H = i \begin{pmatrix} 0 & 1 \\ D_1^2 - m^2 & 0 \end{pmatrix} + \begin{pmatrix} e A_0 & 0 \\ 0 & e A_0 \end{pmatrix} = H_0 + H_1.
\eeq
This operator is hermitean \wrt the inner product
\beq
\label{eq:InnerProduct}
 \braket{\Psi_1}{\Psi_2} = i \int \ud z \left( \phi_1^* \pi_2^* - \pi_1 \phi_2 \right).
\eeq

In the case of Dirichlet boundary conditions, $\Psi(0) = \Psi(1) = 0$, a basis of eigenvectors $\phi^\rD_n$ of $H_0$ with eigenvalues $\omega^\rD_n$ is given by
\begin{align}
 \omega^\rD_n & = \sign(n) \sqrt{m^2 + \pi^2 n^2}, &
 \phi^\rD_n & = \betrag{\omega^\rD_n}^{-\frac{1}{2}} \sin \pi n z,
\end{align}
for $n \in \Z \setminus \{ 0 \}$ and with $\pi_n^* = - i \omega_n \phi_n$. For Neumann boundary conditions $\Psi'(0) = \Psi'(1) = 0$, one has
\begin{align}
 \omega^\rN_{n \neq 0} & = \sign(n) \sqrt{m^2 + \pi^2 n^2}, &
 \phi^\rN_{n \neq 0} & = \betrag{\omega^\rN_n}^{-\frac{1}{2}} \cos \pi n z, \\ 
\label{eq:NeumannMode_0}
 \omega^\rN_{\pm 0} & = \pm m, &
  \phi^\rN_{\pm 0} & = (2m)^{-\frac{1}{2}}.
\end{align}
Note that the modes are normalized to $\sign(n)$, with $\sign(\pm 0) = \pm 1$, and that in the massless limit, the two zero modes $\phi^\rN_{\pm 0}$ of the Neumann case are not normalizable. Physically, this implies the absence of a vacuum in that case.\footnote{In the massless Neumann case, the zero modes correspond to the two-parameter family of solutions $\phi(t, z) = x_0 + t v_0$. It is thus equivalent to a free particle on the line, which quantum mechanically does not possess a ground state, not even a stationary one.}

First order perturbation theory in an indefinite inner product space proceeds analogously to that on Hilbert space. Given a basis $\Psi^{(0)}_n$ of eigenvectors of $H_0$ with non-degenerate eigenvalues $\omega^{(0)}_n$, the first order corrections are given by
\begin{align}
 \omega^{(1)}_n & = \frac{\braket{\Psi_n^{(0)}}{H_1 \Psi_n^{(0)}}}{\braket{\Psi_n^{(0)}}{\Psi_n^{(0)}}}, &
 \Phi^{(1)}_n & = \sum_{k \neq n} \frac{1}{\braket{\Psi^{(0)}_k}{\Psi^{(0)}_k}} \frac{\braket{\Psi^{(0)}_k}{H_1 \Psi^{(0)}_n}}{\omega^{(0)}_n - \omega^{(0)}_k} \Psi^{(0)}_k.
\end{align}
Note however that this breaks down in the presence of an eigenvector $\Psi^{(0)}_k$ of vanishing norm, $\braket{\Psi^{(0)}_k}{\Psi^{(0)}_k} = 0$. In particular, this implies that perturbation theory can not be applied to the massless case with Neumann boundary conditions. For the mode solutions for Dirichlet and Neumann boundary conditions, one finds that there are no corrections to the frequencies $\omega_n$, while for the solutions one obtains, after a lengthy but straightforward calculation,
\begin{align}
\label{eq:DirichletModeExpansion}
 \phi^\rD_n & = \left( m^2 + \pi^2 n^2 \right)^{-\frac{1}{4}} \left[ \sin \pi n z  + \lambda \tfrac{\sqrt{m^2 + \pi^2 n^2}}{2 \pi \betrag{n}} \left( \tfrac{1}{\pi n} ( \tfrac{1}{2} - z) \sin \pi n z - z (1-z) \cos \pi n z \right) \right], \\
\label{eq:NeumannModeExpansion_n}
 \phi^\rN_{n \neq 0} & = \left( m^2 + \pi^2 n^2 \right)^{-\frac{1}{4}} \Big[ \cos \pi n z \nn \\
 & \qquad \qquad \qquad + \lambda \tfrac{\sqrt{m^2 + \pi^2 n^2}}{2 \pi \betrag{n}} \left( \tfrac{1}{\pi n} ( \tfrac{1}{2} - z) \cos \pi n z + ( z (1-z) + (\pi n)^{-2} ) \sin \pi n z \right) \Big], \\
\label{eq:NeumannModeExpansion_0}
 \phi^\rN_{\pm 0} & = (2 m)^{-\frac{1}{2}} \mp \lambda \sqrt{2 m} \left( \tfrac{1}{24} - \tfrac{1}{4} z^2 + \tfrac{1}{6} z^3 \right),
\end{align}
up to corrections of $\order(\lambda^2)$.
In the massless limit, the result for Dirichlet boundary conditions coincides with the expression found in \cite{AmbjornWolfram83}, \cf eq.~(3.5) there. Regarding the result for the Neumann case, it seems that the $\pm 0$ modes were neglected in \cite{AmbjornWolfram83}, and that the massless case was considered in spite of the presence of non-normalizable modes. It seems that disregarding the $\pm 0$ modes is the origin of the anti-screening effect found in \cite{AmbjornWolfram83}. Properly including these modes, which are the most affected by the external field, leads to the expected screening behaviour, as discussed below.

\section{Quantization and point-split renormalization}
\label{sec:Quantization}

Quantization based on the normalized mode solutions discussed in the previous section proceeds by writing the quantum field as
\beq
\label{eq:QuantumField}
 \phi(t,z) = \sum_{n > 0} a_n \phi_n(z) e^{- i \omega_n t} + \sum_{n < 0} b_n^\dag \phi_n(z) e^{- i \omega_n t},
\eeq
with the operators $a_n$, $b_n$, fulfilling the commutation relations
\begin{align}
 [a_n, a_m^\dag] & = \delta_{n m}, &
 [b_n, b_m^\dag] & = \delta_{n m},
\end{align}
with all other commutators vanishing. In the case of Neumann boundary conditions, the $+0$ mode is included in the first sum in \eqref{eq:QuantumField}, while the $-0$ mode is included in the second. The vacuum state $\ket{0}$ is defined by the property that it is annihilated by $a_n$ and $b_n$.

The point-wise products appearing in the charge density are ill-defined and have to be renormalized. A well-controlled way to do this is by point-split renormalization \wrt a Hadamard parametrix. Physically reasonable states, vacuum states in particular \cite{Wrochna}, have two-point functions
\beq
 w^{\phi \phi^*}(x,x') = \bra{\Omega} \phi(x) \phi^*(x') \ket{\Omega}
\eeq 
of Hadamard form, i.e., their singular behavior as $x' \to x$ is of a universal form, which is determined entirely by the background fields in a neighborhood of $x$, but is independent of the state $\Omega$, \cf \cite{Current}, Section 2, for a review. For a charged scalar field in 1+1 dimension, it is of the form
\beq
\label{eq:HadamardForm}
 w^{\phi \phi^*}_{\Omega}(x,x') = - \tfrac{1}{4 \pi} U(x, x') \log (- (x-x')^2 + i \eps (x-x')^0) + R^{\phi \phi^*}_{\Omega}(x,x'),
\eeq
with $U(x,x')$ and $R^{\phi \phi^*}_{\Omega}(x,x')$ smooth functions. While $R^{\phi \phi^*}_{\Omega}(x,x')$ is state dependent, $U(x,x')$ is fixed and given by the parallel transport \wrt the covariant derivative \eqref{eq:D} along the straight line from $x'$ to $x$, up to corrections that are irrelevant for the determination of the vacuum polarization,
\beq
 U(x,x') = \exp \left[ - i e \int_0^1 A_\mu(x' + s(x - x')) (x-x')^\mu \ud s  \right] + \order((x-x')^2).
\eeq
The two point function
\beq
 w^{\phi^* \phi}_{\Omega}(x,x') = \bra{\Omega} \phi^*(x) \phi(x') \ket{\Omega}
\eeq 
has the same form, with $U(x,x')$ replaced by $U(x,x')^* = U(x',x)$. The idea of Hadamard point-split renormalization, which goes back to Dirac \cite{Dirac34} and was rediscovered in the context of quantum field theory on curved space-times, \cf \cite{HollandsWald} for a recent review, is to define the expectation value of a local expression quadratic in fields by
\beq
\label{eq:HadamardPointSplit}
 \bra{\Omega} D_\alpha \phi(x) (D_\beta \phi)^*(x) \ket{\Omega} = \lim_{x' \to x} \left[ D_\alpha D'^*_{\beta'} \left( w^{\phi \phi^*}_{\Omega}(x,x') - H^{\phi \phi^*}(x,x') \right) \right].
\eeq
Here $\alpha$ and $\beta$ are symmetrized multiindices, $D'^*_{\mu}$ stands for the application of $D^*_\mu = \del_\mu - i e A_\mu$ on the primed variable, and $H^{\phi \phi^*}(x,x')$ is the first term on the \rhs of \eqref{eq:HadamardForm}.

For the evaluation of the vacuum polarization, we have two such expressions to evaluate. We perform the point-splitting in the time direction, so that $x' = (t + \tau, z)$, and we obtain
\begin{align}
 \bra{0} \phi^*(x) D_0 \phi(x') \ket{0} & = - i \sum_{n < 0} \betrag{\phi_n(z)}^2 (\omega_n - e A_0) e^{- i \omega_n (\tau+i \eps)}, \\
 \bra{0} \phi(x) D^*_0 \phi^*(x') \ket{0} & = i \sum_{n > 0} \betrag{\phi_n(z)}^2 (\omega_n - e A_0) e^{i \omega_n (\tau+i \eps)},
\end{align}
where we added an $i \eps$ prescription to ensure convergence. Note that these expressions are valid generically, i.e., not only for the external field given by \eqref{eq:A} (of course the modes $\phi_n$ differ for different external fields). On the other hand, we compute
\begin{align}
\label{eq:Parametrix_1}
 D'_0 H^{\phi^* \phi}(x,x') & = - \frac{1}{2\pi} \frac{1}{\tau + i \eps} U(x', x) + \order(\tau) = - \frac{1}{2\pi} \left( \frac{1}{\tau + i \eps} - i e A_0 \right) + \order(\tau), \\
\label{eq:Parametrix_2}
 D'^*_0 H^{\phi \phi^*}(x,x') & = - \frac{1}{2\pi} \frac{1}{\tau + i \eps} U(x, x') + \order(\tau) = - \frac{1}{2\pi} \left( \frac{1}{\tau + i \eps} + i e A_0 \right) + \order(\tau).
\end{align}
For the vacuum polarization, we thus obtain
\begin{align}
\label{eq:VacuumPolarization}
 \rho(z) & = i  e \bra{0} \left( \phi^* D_0 \phi - \phi D_0^* \phi^* \right) \ket{0} \\
 & = e \lim_{\tau \to 0} \left( \sum_{n<0} \betrag{\phi_n(z)}^2 (\omega_n - e A_0) e^{- i \omega_n (\tau+i \eps)} + \sum_{n > 0} \betrag{\phi_n(z)}^2 (\omega_n - e A_0) e^{i \omega_n (\tau+i \eps)} \right) + \frac{e^2}{\pi} A_0(z). \nn
\end{align}
We see that the singular parts of \eqref{eq:Parametrix_1} and \eqref{eq:Parametrix_2} cancel, and similarly, the singularities in the coinciding point limit from the two sums must cancel. But to obtain the correct finite result, some care has to be taken. If the summation and the limit $\tau \to 0$ in \eqref{eq:VacuumPolarization} were interchangeable, one would obtain\footnote{Here the addition ``(WRONG)'' is meant to indicate that the equation is not correct in general, but only under the mentioned assumption, namely interchangeability of the limit and the summation.}
\beq
\label{eq:rhoWRONG}
  \rho(z) = e\sum_{n > 0} \left( \betrag{\phi_n(z)}^2 (\omega_n - e A_0) - \betrag{\phi_{-n}(z)}^2 (\omega_n + e A_0) \right) + \frac{1}{\pi} e^2 A_0(z) \qquad \text{(WRONG)}.
\eeq
This expression, however, is not gauge invariant (the sum over $n$ is, but the second term is not), while \eqref{eq:VacuumPolarization}, when evaluated in the correct order, is explicitly gauge invariant. This implies that the summation and the limit $\tau \to 0$ are in general not interchangeable. They may be in special cases, but then this is gauge dependent. In the case of a discrete spectrum that we are interested in here, the interchange might be possible if $\omega_{-n} = - \omega_n$ (which is a gauge dependent statement) so that the two sums in \eqref{eq:VacuumPolarization} can be combined to a single sum over $n > 0$, involving a single oscillatory factor $e^{i \omega_n (\tau + i \eps)}$. If its coefficient decays quickly enough in $n$, then the interchange is possible. One then does obtain \eqref{eq:rhoWRONG}, which coincides with the expression used in \cite{AmbjornWolfram83}, up to the last term, which came from the Hadamard point-split procedure. 

It may be instructive to provide an alternative derivation of \eqref{eq:VacuumPolarization}. We have seen that the divergences of the coinciding point limit cancel out in the charge density, if we consider
\beq
\label{eq:CurrentDifference}
 \bra{0} \phi^*(x) D_0 \phi(x') \ket{0} - \bra{0} \phi(x) D^*_0 \phi^*(x') \ket{0},
\eeq 
where still $x' = (t+ \tau, z)$. One may thus be tempted to assume that the limit $\tau \to 0$ commutes with the summation over modes and take this as the definition of the vacuum charge density. One then obtains \eqref{eq:rhoWRONG}, without the last term, i.e., the expression for the vacuum charge density that was used in \cite{AmbjornWolfram83}. However, this approach has the following deficiency: One can interpret the point-splitting as a regularization. However, to ensure that final renormalized result is gauge invariant, the regularization should be gauge invariant, too. But the expression \eqref{eq:CurrentDifference} is not gauge invariant for $x \neq x'$.\footnote{The same could be said about the expression \eqref{eq:HadamardPointSplit}, which underlies the Hadamard point-split renormalization. However, unlike the terms in \eqref{eq:CurrentDifference}, the expression in square brackets in \eqref{eq:HadamardPointSplit} is smooth, so one may safely take the limit of coinciding points to obtain a manifestly gauge invariant result. One could of course also equip \eqref{eq:HadamardPointSplit} with a parallel transport to ensure gauge invariance for $x \neq x'$, but this would not affect the coinciding point limit, as $U(x,x) = 1$.} Instead, one should consider
\beq
\label{eq:PointSplitGaugeInvariant}
 \bra{0} \phi^*(x) D_0 \phi(x') \ket{0} U(x,x') - \bra{0} \phi(x) D^*_0 \phi^*(x') \ket{0} U(x',x),
\eeq 
which is gauge invariant. However, recalling that $U(x,x') = 1 + i e A_0(z) \tau + \order(\tau^2)$ and that the ground state is Hadamard, so that
\begin{align}
 \bra{0} \phi^*(x) D_0 \phi(x') \ket{0} & = - \frac{1}{2 \pi} \frac{1}{\tau + i \eps} + \order(\tau^0), &
 \bra{0} \phi(x) D^*_0 \phi^*(x') \ket{0} & = - \frac{1}{2 \pi} \frac{1}{\tau + i \eps} +  \order(\tau^0),
\end{align}
\cf \eqref{eq:Parametrix_1}, \eqref{eq:Parametrix_2}, we may rewrite the coinciding point limit of \eqref{eq:PointSplitGaugeInvariant} as
\begin{multline}
 \lim_{\tau \to 0} \left( \bra{0} \phi^*(x) D_0 \phi(x') \ket{0} U(x,x') - \bra{0} \phi(x) D^*_0 \phi^*(x') \ket{0} U(x',x) \right) \\
 = \lim_{\tau \to 0} \left( \bra{0} \phi^*(x) D_0 \phi(x') \ket{0} - \frac{i e}{2\pi} A_0(z) - \bra{0} \phi(x) D^*_0 \phi^*(x') \ket{0} - \frac{i e}{2\pi} A_0(z) \right).
\end{multline}
Multiplication by $i e$ yields precisely our expression \eqref{eq:VacuumPolarization} for the vacuum polarization. Hence, our expression for the vacuum polarization can also be derived without using a Hadamard point-split, but by insisting on a gauge invariant regularization via point-splitting.

As an historical aside, the necessity of introducing a parallel transport in a point-split regularization of the current was first pointed out by Schwinger \cite{Schwinger:1959xd} in the context of Dirac fields. This note was published several years after Schwinger's seminal paper \cite{Schwinger:1951nm} on the vacuum polarization, in which the point-splitting was performed without the parallel transport. In the meantime, Wichmann and Kroll \cite{WichmannKroll} used Schwinger's original point-splitting prescription (without inclusion of the parallel transport) to derive the mode sum formula for the vacuum polarization in the context of Dirac fields in 3+1 dimensions. Their expression suffers from the same deficiency as the mode sum formula used in \cite{AmbjornWolfram83}: To restore gauge invariance, one has to add a correction term, $\frac{e}{3 \pi^2} A_0^3$, to the mode sum, \cf \cite{SoffMohr88} for example. This is precisely the 3+1 dimensional analog of the second term on the \rhs of \eqref{eq:VacuumPolarization} and \eqref{eq:rhoWRONG}. We refer to Section~2.2 of \cite{Current} for further (historical) comments on the intricacies of point-split renormalization of the vacuum polarization.

\section{Perturbative evaluation of the vacuum polarization}

\begin{figure}
\centering
\includegraphics[width=0.7 \textwidth]{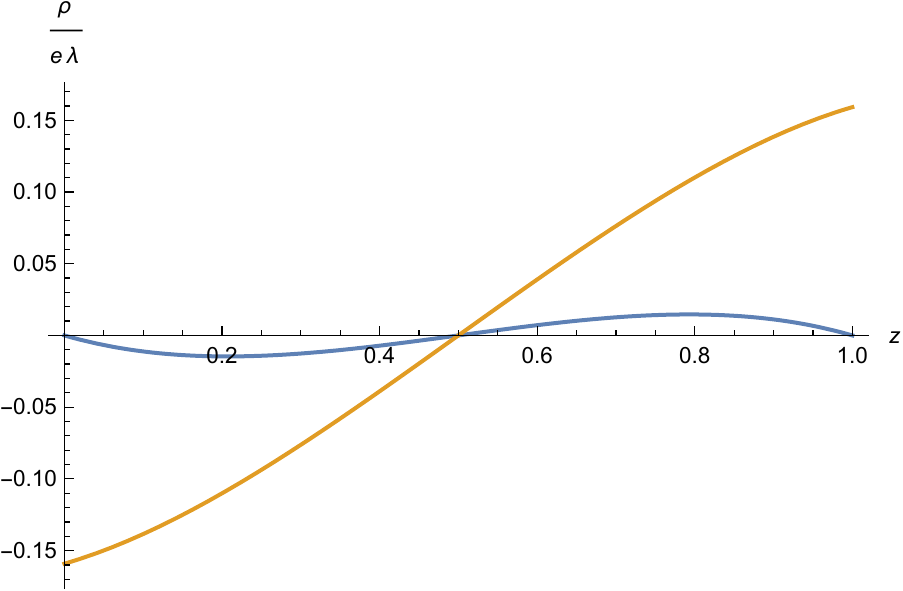}
\caption{In blue the vacuum polarization for Dirichlet boundary conditions in the massless case, at first order in $\lambda$, according to \eqref{eq:rho_DirichletMassless}. In orange the result of \cite{AmbjornWolfram83}.}
\label{fig:DirichletMassless}
\end{figure}

We now want to evaluate the expression \eqref{eq:VacuumPolarization}. In the perturbative approach, at first order in $\lambda$, this is possible analytically in the massless case with Dirichlet boundary conditions. We rewrite \eqref{eq:VacuumPolarization} as
\beq
 \rho(z) = e \lim_{\tau \to 0} \sum_{n = 1}^\infty \left( \betrag{\phi_n(z)}^2 (\pi n - e A_0) - \betrag{\phi_{-n}(z)}^2 (\pi n + e A_0) \right) e^{i \pi n (\tau + i \eps)} + \frac{1}{\pi} e^2 A_0(z).
\eeq
Up to first order in $\lambda$, we have
\beq
 \betrag{\phi_n(z)}^2 = \frac{1}{\pi \betrag{n}} \left( \sin^2 \pi n z - \lambda \sin \pi n z \left[ \frac{1}{\pi n} \left( z - \frac{1}{2} \right) \sin \pi n z + z(1-z) \cos \pi n z \right] \right)
\eeq
and thus obtain, to first order in $\lambda$,
\begin{align}
 \rho(z) & = - 2 e \lambda \lim_{\tau \to 0} \sum_{n=1}^\infty z(1-z) \sin \pi n z \cos \pi n z e^{i \pi n (\tau + i \eps)} - \frac{1}{\pi} e \lambda \left( z - \frac{1}{2} \right) \nn \\
 & = - e \lambda z (1-z) \lim_{\tau \to 0} \frac{1}{2i} \left( \frac{e^{i \pi (2 z + \tau + i \eps)}}{1 - e^{i \pi (2 z + \tau + i \eps)}} - \frac{e^{i \pi (-2 z + \tau + i \eps)}}{1 - e^{i \pi (-2 z + \tau + i \eps)}} \right) - \frac{1}{\pi} e \lambda \left( z - \frac{1}{2} \right) \nn \\
\label{eq:rho_DirichletMassless}
 & = - e \lambda \frac{z (1-z)}{2} \cot \pi z - \frac{1}{\pi} e \lambda \left( z - \frac{1}{2} \right).
\end{align}
Up to the last term, which is precisely the last term in \eqref{eq:VacuumPolarization}, this coincides with the result of \cite{AmbjornWolfram83}.\footnote{Note that without the point-split with $i \eps$ prescription the sum in the first line of \eqref{eq:rho_DirichletMassless} is not convergent. Hence, to obtain the result without this prescription, as in \cite{AmbjornWolfram83}, a damping factor $e^{- n \eps}$ has to be introduced by hand. In this sense, the interchange of the limit $\tau \to 0$ and the summation in \eqref{eq:VacuumPolarization} is not permissible, even in our gauge where $\omega_{-n} = - \omega_n$.} However, this term makes a qualitative difference for the resulting vacuum polarization, as seen in Figure~\ref{fig:DirichletMassless}. The vacuum polarization charge density is not only much smaller in magnitude, but it vanishes exactly at the boundaries, as one might naively expect for Dirichlet, i.e., repulsive, boundary conditions.

\begin{figure}
\centering
\includegraphics[width=0.7 \textwidth]{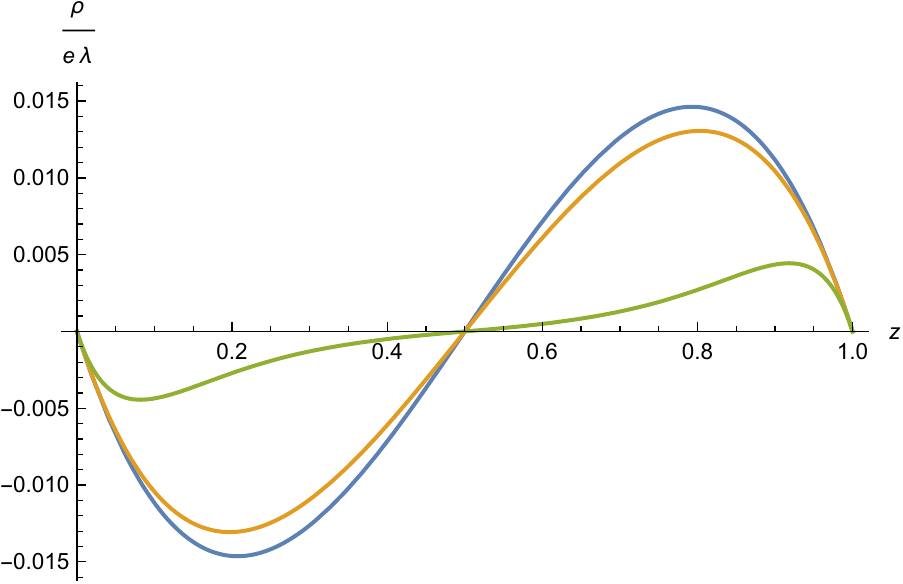}
\caption{Vacuum polarization for Dirichlet boundary conditions to first order in $\lambda$ for $m = 0$ (blue), $m = 1$ (orange), and $m = 5$ (green).}
\label{fig:Dirichlet}
\end{figure}

In the massive case, a completely analytic treatment is not possible. However, in the perturbative treatment, the numerical evaluation of the expression \eqref{eq:VacuumPolarization}, using the perturbative expansion \eqref{eq:DirichletModeExpansion} of the mode functions, is straightforward. In order to get rid of the limit $\tau \to 0$ in \eqref{eq:VacuumPolarization}, one can use the following trick: One subtracts and adds the expression
\beq
\label{eq:TrickTerm}
 - 2 e \lambda \lim_{\tau \to 0} \sum_{n=1}^\infty z (1-z) \sin \pi n z \cos \pi n z \ e^{i \pi n (\tau + i \eps)},
\eeq
which was obtained in the perturbative treatment of the massless case, \cf \eqref{eq:rho_DirichletMassless}. The subtracted term is included into the mode sum \eqref{eq:VacuumPolarization} and improves its convergence, so that the limit $\tau \to 0$ can be safely performed and the summation cut off at some large $N$ (in the range of masses considered here, $N = 50$ is sufficient). The added term \eqref{eq:TrickTerm} can be handled as in \eqref{eq:rho_DirichletMassless}, i.e., replaced by $- e \lambda \frac{z (1-z)}{2} \cot \pi z$.
 The results thus obtained for the Dirichlet case are shown in Figure~\ref{fig:Dirichlet}. We see that vacuum polarization is suppressed for higher masses.

\begin{figure}
\centering
\includegraphics[width=0.7 \textwidth]{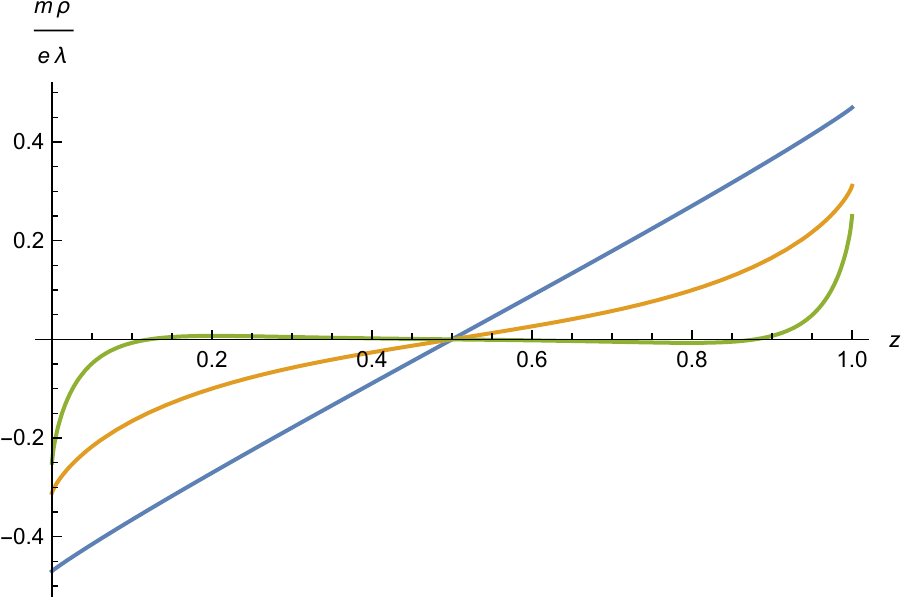}
\caption{Vacuum polarization for Neumann boundary conditions to first order in $\lambda$ for $m = 0.1$ (blue), $m = 1$ (orange), and $m = 5$ (green). Note that the charge density is multiplied by $m$ in this plot to achieve better visibility.}
\label{fig:Neumann}
\end{figure}

Perturbative results for the case of Neumann boundary conditions are shown in Figure~\ref{fig:Neumann}. These are obtained analogously to the numerical results for Dirichlet boundary conditions, i.e., by subtracting and adding \eqref{eq:TrickTerm} and using the procedure described above. Again, we see that vacuum polarization is suppressed for increasing mass (note that $m \rho$ is plotted). The results also indicate that the vacuum polarization diverges as $m \to 0$, which is a consequence of the divergence of the $\pm 0$ mode \eqref{eq:NeumannModeExpansion_0} in that limit. This divergence is further discussed in the next section in the context of a non-perturbative analysis. Another notable result is that, in contrast to the Dirichlet case, the vacuum polarization is not vanishing at the boundary, it is in fact maximal there, as one would expect for attractive boundary conditions. In any case, it is screening, as opposed to the anti-screening behaviour that was found in \cite{AmbjornWolfram83}. As discussed above, it is the proper inclusion of the $\pm 0$ modes, which also necessitated to work with a finite mass, which explains the difference to the results of \cite{AmbjornWolfram83} (apart from the inclusion of the last term on the \rhs of \eqref{eq:VacuumPolarization}). Finally, we remark that for Neumann boundary conditions and high enough masses, the vacuum polarization changes sign within the interval $(0, \frac{1}{2})$, and analogously in $(\frac{1}{2}, 1)$. This means that, in the present approximation, as one approaches the boundary at $z=0$ from $z=\frac{1}{2}$, the perceived charge at $z=0$ first decreases, and then increases. Such a somewhat counterintuitive behaviour is also present in the vacuum polarization in the Coulomb potential \cite{SoffMohr88}, although there it occurs upon including effects of higher order in the external field.

\section{Non-perturbative evaluation of the vacuum polarization}

In order to verify the validity of the perturbative approximation, and to gain some insight into the nature of the divergence as $m \to 0$ seen in the perturbative results, we now perform a non-perturbative treatment, based on the exact mode solutions \eqref{eq:ExactModeSolution}. One first numerically determines the frequencies $\omega_n$ solving the desired boundary conditions and then normalizes according to \eqref{eq:InnerProduct}. One can then use the same trick as above, i.e., subtracting and adding \eqref{eq:TrickTerm}. However, it turns out that the result obtained for a cut-off of the mode sum at $N$ still oscillates, as a function of $z$, with a period $\Delta z = \frac{1}{N+1}$. The cut-off dependence shows that this is clearly an artefact of the cut-off. The oscillations are thus removed by a suitable averaging.

\begin{figure}
\centering
\includegraphics[width=0.7 \textwidth]{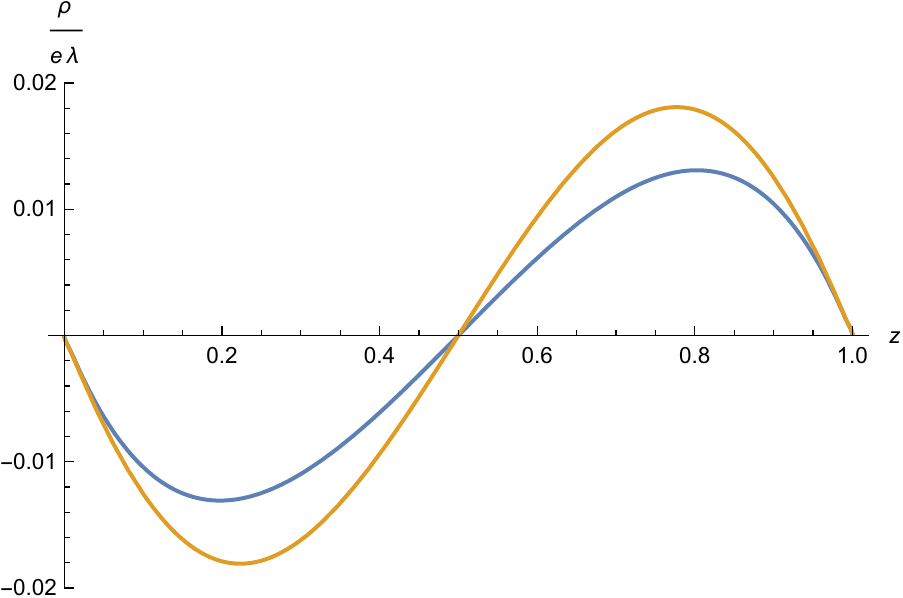}
\caption{Vacuum polarization for Dirichlet boundary conditions, non perturbative in $\lambda$, for $m = 1$, with $\lambda = 1$ (blue), and $\lambda = 10$ (orange). Note that for better comparison, the charge density is divided by $\lambda$.}
\label{fig:DirichletNP}
\end{figure}

The results thus obtained for Dirichlet boundary conditions are shown in Figure~\ref{fig:DirichletNP}. Not shown in the plot is the result for the perturbative calculation for the same parameters, as it is indistinguishable from the result for $\lambda = 1$. This shows that for $\lambda \lesssim 1$, the perturbative result is a very good approximation. For $\lambda = 10$, deviations from the perturbative result are clearly visible, but not huge. Investigating the non-perturbative results for different masses, one finds, perhaps not surprisingly, that, for fixed $\lambda$, the deviations from the perturbative results are more pronounced the smaller the mass is.

\begin{figure}
\centering
\includegraphics[width=0.7 \textwidth]{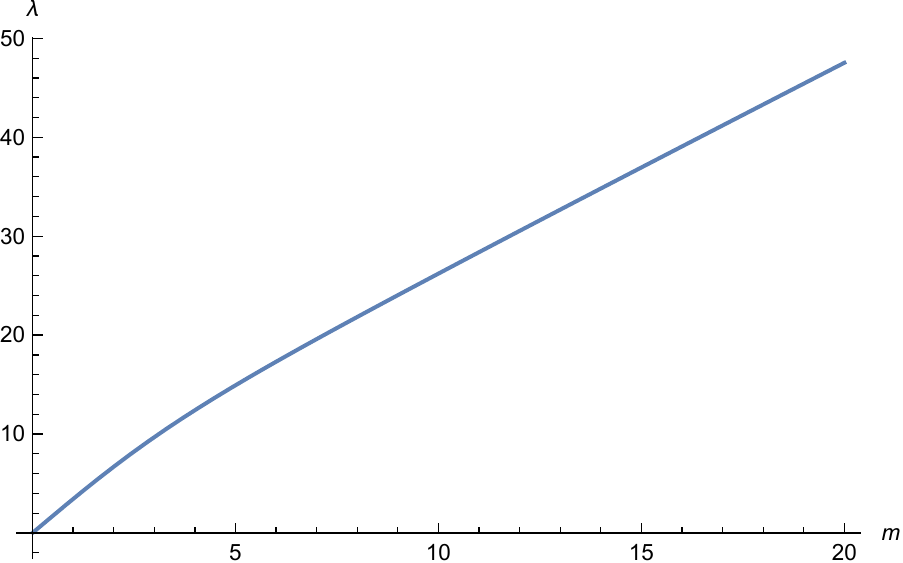}
\caption{The critical value of $\lambda$ as a function of the mass $m$. In the parameter region above this line, the system exhibits an instability (existence of imaginary frequency modes) for Neumann boundary conditions.}
\label{fig:Instability}
\end{figure}

Finally, we study the case of Neumann boundary conditions. We make the following crucial observation: Given a finite mass $m > 0$ and turning on the external field, i.e., increasing $\lambda$ from $0$, the eigenvalue $\omega_{+ 0}$ of the lowest positive frequency mode decreases, until it reaches $0$ at a finite value of $\lambda$. At this critical value, the mode ceases to be normalizable \wrt the inner product \eqref{eq:InnerProduct}, analogously to the zero mode \eqref{eq:NeumannMode_0} in the massless case at $\lambda = 0$ (and indeed the critical value of $\lambda$ for $m = 0$ is $\lambda = 0$). Further increasing $\lambda$, the eigenvalues $\omega_{\pm 0}$ move to the imaginary axis. Physically, this implies an instability (exponential growth instead of oscillations). In Figure~\ref{fig:Instability}, we plot the critical value of $\lambda$ against the mass $m$.

\begin{figure}
\centering
\includegraphics[width=0.7 \textwidth]{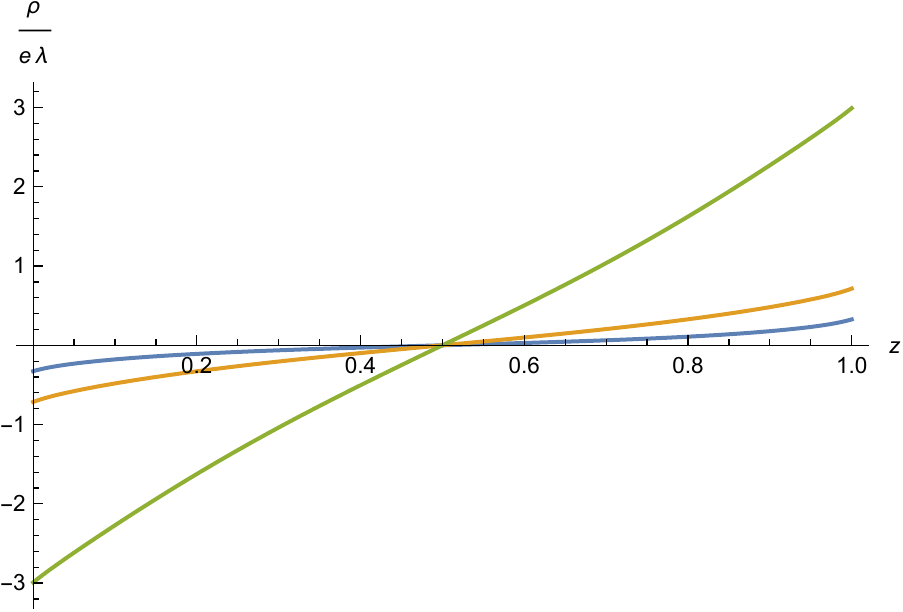}
\caption{Vacuum polarization for Neumann boundary conditions for $m = 1$ and $\lambda = 1$ (blue), $\lambda = 3$ (orange), and $\lambda = 3.4$ (green). The critical value of $\lambda$ for the given $m$ is $\lambda = 3.432$.}
\label{fig:NeumannNP}
\end{figure}

For the non-perturbative evaluation of the vacuum polarization, we should thus restrict to sub-critical field strength. Figure~\ref{fig:NeumannNP} shows results for different values of $\lambda$ for fixed mass $m$. While the result for $\lambda = 1$ is barely distinguishable from the perturbative result (not shown in Figure~\ref{fig:NeumannNP}), we see that the vacuum polarization becomes strongly non-linear in $\lambda$ and in fact diverges as one approaches criticality. This divergence is analogous to the divergence in the limit $m \to 0$ observed in the perturbative calculation, and it has the same origin: As one approaches criticality, the modes $\phi_{\pm 0}$ diverge (recall the behaviour of \eqref{eq:NeumannModeExpansion_0} as $m \to 0$).

\subsection*{Acknowledgements}

J.Z.~thanks 
S.~Erb and M.~Wussing for their help in re-establishing contact with J.W, and 
J.~Ambj{\o}rn for discussions on the point-split renormalization of the vacuum polarization, which were helpful for sharpening the arguments put forward at the end of Section~\ref{sec:Quantization}.



\end{document}